# Cell-Probe Lower Bounds for Prefix Sums

Emanuele Viola[*]

October 30, 2018

**Abstract**

We prove that to store $n$ bits $x \in \{0,1\}^n$ so that each prefix sum (a.k.a. rank) query $\text{Sum}(i) := \sum_{k \le i} x_k$ can be answered by non-adaptively probing $q$ cells of $\lg n$ bits, one needs memory

$$n + n/\log^{O(q)} n.$$

This matches a recent upper bound of $n + n/\log^{\Omega(q)} n$ by Pǎtraşcu (FOCS 2008), also non-adaptive.

We also obtain a $n + n/\log^{2^{O(q)}} n$ lower bound for storing a string of balanced brackets so that each Match($i$) query can be answered by non-adaptively probing $q$ cells. To obtain these bounds we show that a too efficient data structure allows us to break the correlations between query answers.

[*]Supported by NSF grant CCF-0845003. Email: `viola@ccs.neu.edu`

# 1 Introduction

The problem of succinctly storing $n$ bits $x \in \{0,1\}^n$ so that each prefix sum (a.k.a. rank) query $\text{Sum}(i) := \sum_{k \leq i} x_k$ can be answered efficiently is a fundamental data structure problem that has been studied for more than two decades. The best known upper bound for this problem is a data structure by Pătraşcu which answers queries by probing $q$ cells of $\lg n$ bits, and uses memory

$$n + n/\lg^{\Omega(q)} n, \qquad (1)$$

see [Păt08] and the references there.

We prove the first lower bound for this problem, matching Pătraşcu's upper bound (1) for non-adaptive schemes. We remark that known upper bounds, including [Păt08] in Equation (1), are also non-adaptive.

**Theorem 1.1** (Lower bound for prefix sums). *To store $\{0,1\}^n$ in $[n]^u$ so that each $\text{Sum}(i) := \sum_{k \leq i} x_k$ query can be computed by non-adaptively probing $q$ cells of $\lg_2 n$ bits, one needs memory*

$$u \cdot \lg_2 n \geq n - 1 + n/\lg^{A \cdot q} n,$$

*where $A$ is an absolute constant.*

Our techniques apply to other problems as well; for example we obtain the following lower bound for the problem of storing strings of balanced brackets so that indexes to matching brackets can be retrieved quickly.

**Theorem 1.2** (Lower bound for balanced brackets). *To store $\text{Bal} := \{x \in \{0,1\}^n : x \text{ corresponds to a string of balanced brackets}\}$ in $[n]^u$, $n$ even, so that each $\text{Match}(i)$ query can be computed by non-adaptively probing $q$ cells of $\lg_2 n$ bits, one needs memory*

$$u \cdot \lg_2 n \geq n - 1 + n/\lg^{A^q} n,$$

*where $A$ is an absolute constant.*

The best known upper bound for this problem is again $n + n/\lg^{\Omega(q)} n$, non-adaptive [Păt08]. It is an interesting open problem to close the gap between that and our lower bound of $n + n/\lg^{2^{O(q)}} n$.

## 1.1 Techniques

We now explain the techniques we use to prove our lower bound for prefix sums. We show that a too efficient data structure would allow us to *break the dependencies* between the various prefix sums and obtain the following contradiction: For three indexes $1 \leq p < i < j \leq n$, a subset $X \subseteq \{0,1\}^n$ of inputs, and some integers $s, s'$:

$$1/1000 \geq \Pr_{x \in X}\left[\sum_{k \leq j} x_k \geq s \wedge \sum_{k \leq i} x_k < s'\right] \qquad (2)$$

$$\approx \Pr_{x \in X}\left[\sum_{k \leq j} x_k \geq s\right] \cdot \Pr_{x \in X}\left[\sum_{k \leq i} x_k < s'\right] \qquad (3)$$

$$\geq \frac{1}{10} \cdot \frac{1}{10} \gg 1/1000. \qquad (4)$$

We now explain how we obtain these inequalities.

**Approximation (3):** For the Approximation (3) we ignore the integers $s, s'$ and more generally show that the distributions of the sums $\sum_{k \leq i} x_k$ and $\sum_{k \leq j} x_k$ are statistically close to independent. By the data structure, for every input $x \in \{0,1\}^n$ and index $i$ the sum $\sum_{k \leq i} x_k$ can be retrieved as a function $d_i$ of $q$ cells $Q(i) \subseteq [u]$ of the encoding $Enc(x)$ of $x$:

$$\sum_{k \leq i} x_k = d_i\left(Enc(x)|_{Q(i)}\right),$$

where $Enc(x)|_{Q(i)}$ denotes the $q$ cells of size $n$ ($\lg n$ bits) of $Enc(x) \in [n]^u$ indexed by $Q(i)$.

An evident case in which the two sums $\sum_{k \leq i} x_k$ and $\sum_{k \leq j} x_k$ could correlate (for a random input $x \in \{0,1\}^n$) is when $Q(i) \cap Q(j) \neq \emptyset$. To avoid this, we prove a separator Lemma 2.3 yielding a set $B$ of size $n \lg^b n$ such that there are $n/\lg^a n \gg |B|$ disjoint sets among

$$Q(1) \setminus B, Q(2) \setminus B, \ldots, Q(n) \setminus B.$$

We denote by $V \subseteq [n]$ the set of indices of the disjoint sets.

The proof of the separator lemma is an inductive argument based on coverings; it resembles arguments appearing in other contexts (cf. [Vio07, Overview of the Proof of Lemma 11]).

Via an averaging argument we fix the values of the cells whose index $\in B$ so as to have a set of inputs $X \subseteq \{0,1\}^n$ of size $|X| \geq 2^n/[n]^{|B|}$ such that the $n/\lg^a n$ sums $\sum_{k \leq i} x_k$ for $i \in V$ can be recovered by reading disjoint cells, which we again denote $Q(i)$ (so now we have $Q(i) \cap Q(j) = \emptyset$ for $i, j \in V$).

This takes care of the "evident" problem that different prefix sums may be answered by reading overlapping sets of cells. Of course, there could be other types of correlations arising from the particular distribution that a random input $x \in X$ induces in the values of the cells. To handle these, we rely on a by-now standard information-theoretic lemma that guarantees the existence of a set of cell indices $G \subseteq [u]$ such that any $2q$ cells whose indices $\in G$ are jointly uniformly distributed (cf. [Vio09, §2] and the references therein). Thus, if we find $i, j \in V$ such that $Q(i) \cup Q(j) \subseteq G$ we conclude the derivation of Approximation

(3) as follows

$$\left(\sum_{k\leq i} x_k, \sum_{k\leq j} x_k\right)_{x\in X} = \left(d_i\left(Enc(x)|_{Q(i)}\right), d_j\left(Enc(x)|_{Q(j)}\right)\right)_{x\in X} \approx$$

$$\left(d_i\left(U|_{Q(i)}\right), d_j\left(U|_{Q(j)}\right)\right)_U = \left(d_i\left(U|_{Q(i)}\right)\right)_U \cdot \left(d_j\left(U|_{Q(j)}\right)\right)_U \approx \left(\sum_{k\leq i} x_k\right)_{x\in X} \cdot \left(\sum_{k\leq j} x_k\right)_{x\in X},$$

where $U$ denotes the uniform distribution over the cell values. The first equality is by definition, the approximations are by the information-theoretic lemma mentioned above, and the second equality holds because $Q(i) \cap Q(j) = \emptyset$.

To conclude this part, it only remains to see that there are indeed $i, j \in V$ such that $Q(i) \cup Q(j) \subseteq G$. The size of $G$ is related to the redundancy $r = n/\lg^{A \cdot q} n$ of the data structure, specifically $|[u] \setminus G| = O(q \cdot r)$. As the sets $Q(i)$ for $i \in V$ are disjoint, we have a set $V'$ of indices of size at least $|V'| \geq |V| - |[u] \setminus G| \geq n/\lg^a n - q \cdot r \geq \Omega(n/\lg^a n)$ such that Approximation (3) is satisfied by any $i, j \in V'$.

**Inequalities (2) and (4).** To prove Inequalities (2) and (4) we reason in two steps. First, we find three indices $p < i < j$, where $i$ and $j$ are both in $V'$, such that the entropy of the $j - p$ variables $x_{p+1} \ldots x_j$, for $x \in X$, is large *even when conditioned on all the others before them*, and moreover $i - p \geq c(j - i)$ for a large constant $c$. Specifically, we have the picture

$$x_1 x_2 \underline{\qquad\qquad} x_p \underbrace{x_{p+1} x_{p+2} \underline{\qquad\qquad} x_i}_{\geq c \cdot d = c(j-i)} \underbrace{x_{i+1} x_{i+2} \underline{\qquad} x_j}_{d := j-i}$$

and the guarantee that

$$H(x_{p+1}, x_{p+2}, \ldots, x_j | x_1, x_2, \ldots, x_p) \geq (j - p) - \epsilon. \qquad (5)$$

Second, from Equation (5) we obtain the integers $s, s'$ to satisfy Inequality (4).

For the first step we start by selecting a subset of the indices in $V'$ that partitions $[n]$ in intervals such that the first is $\geq c$ times larger than the second, the third is $\geq c$ times larger than the fourth and so on:

$$0\underline{\qquad\qquad}v'_1\underline{\ }v'_2\underline{\qquad\qquad}v'_3\underline{\ }v'_4\underline{\qquad}\ldots\underline{\qquad}v'_{2k}\underline{\qquad\qquad}v'_{2k+1}\underline{\ }v'_{2k+2}\ldots$$

A simple argument shows that we can find a subset of indices as above whose size is a $\Omega(1/\lg n)$ fraction of $|V'|$.

We then view $x \in X$ as the concatenation of random variables $Z_0, Z_1, \ldots$, each spanning two adjacent intervals:

$$\underbrace{x_1\underline{\qquad}x_{v'_1}\underline{\ }x_{v'_2}}_{Z_0} \underbrace{x_{v'_2+1}\underline{\qquad}x_{v'_3}\underline{\ }x_{v'_4}}_{Z_1} \underbrace{x_{v'_4+1}\underline{\qquad}x_{v'_5}\underline{\ }x_{v'_6}}_{Z_2} \ldots$$

We would like to find a $Z_k$ that has high entropy even when conditioned on the previous ones. We make the simple and key observation that known proofs of the information-theoretic lemma (e.g., the one in [SV08]) give this. Specifically, we want to avoid the variables $Z_k$ that have a distribution far from uniform as a result of the entropy we lost when we went from the set of all inputs $\{0,1\}^n$ to $X \subseteq \{0,1\}^n$. In the fixing we lost $\lg n \cdot |B| = n/\lg^{b-1} n$ bits of entropy (since each cell contains $\lg n$ bits), and by an information-theoretic lemma there are only $O(n/\lg^{b-1} n)$ bad variables $Z_k$. Since the number of variables $Z_k$ is $\Omega(|V'|/\lg n) \geq n/\lg^{a+2} n$, by taking $b \geq a+4$ we see that most variables $Z_k$ satisfy Equation (5). Note here we rely on the separator lemma giving us a set $B$ whose removal yields a number of disjoint sets $|V| = n/\lg^a n \gg n/\lg^b n = |B|$.

For the second step, we reason as follows. For simplicity, let us think of a typical value $t$ of $x_1 + x_2 + \ldots + x_p$ for $x \in X$ (we actually pick a $t$ which "cuts in half" the outputs of $x_1 + x_2 + \ldots + x_p$). By Equation (5), we can think of the distribution of $x_{p+1}, x_{p+2}, \ldots, x_j$ as $j - p$ independent coin tosses, and this is true even after we condition on the first $p$ bits summing to $t$. We set the integers $s := t + (j-p)/2 + c^{1/3}\sqrt{d}$, and $s' := t + (i-p)/2$, where recall $d := j - i$.

To see Inequality (2), note that for it to be true is must be the case that the sum of $j - i = d$ coin tosses exceeds its mean by $c^{1/3}\sqrt{d}$, which has probability $\leq 1/1000$ for large enough $c$.

For the Inequalities (4), note that $\Pr_{x \in X}\left[\sum_{k \leq i} x_k < s'\right]$ has probability about $1/2$ as it is just the probability that a sum of coin tosses does not exceed its mean. Finally, $\Pr_{x \in X}\left[\sum_{k \leq j} x_k \geq s\right]$ is the probability that the sum of $j - p \geq c \cdot d$ coin tosses exceeds its mean by $c^{1/3}\sqrt{d} \leq c^{1/3}\sqrt{(j-p)/c} = \sqrt{j-p}/c^{1/6}$; this probability is $\geq 1/10$ for a sufficiently large $c$.

This completes the overview of our proof.

**Comparison with [Vio09].** In this section we compare our techniques with those in [Vio09]. To illustrate the latter, consider the problem of storing $n$ ternary elements $t_1, \ldots, t_n \in \{0, 1, 2\}$ in $u$ bits (not cells) so that each ternary element can be retrieved by reading just $q$ bits. The main idea in [Vio09] is to use the fact that if the data structure is too succinct, then, for a random input, the query answers are a function of $q$ *almost uniform* bits. But the probability that a uniform ternary element in $\{0, 1, 2\}$ is 2 equals $1/3$, whereas the probability that a function of $q$ uniform bits is 2 equals $A/2^q$ for some integer $A$. Since the gap between $1/3$ and $A/2^q$ is $\Omega(2^{-q})$, if the $q$ bits used to retrieve the ternary element are $o(2^{-q})$-close to uniform we reach a contradiction.

This technique cannot be used when reading $q \geq \lg n$ bits, because one would need the bits to be $2^{-q} \leq 1/n$ close to uniform, which cannot be guaranteed (cf. the parameters of Lemma 2.5: the dependence on the error parameter is tight up to constant factors as can be verified by conditioning on the event that the majority of the bits is 1). The same problem arises with non-boolean queries like prefix sums; probing two cells gives $2\lg n$ bits and granularity $1/n^2$ in the probabilities of the query output, which can be designed to be at statistical distance $\leq 1/n$ from the correct distribution.

This work departs from the idea of focusing on the distribution of a single query output, and rather focuses on the correlation between different query outputs.

**Organization:** In §2 we formally state and prove our lower bound for prefix sums relying on three lemmas which are proven in §3. In §4 we prove our lower bound for matching brackets. We conclude in §5 with some open problems.

## 2 Lower bound for prefix sums

We now formally state the data structure problem and then recall our main result.

**Definition 2.1** (Data structure for prefix sums). *We say that we store $\{0,1\}^n$ in $[n]^u$ supporting prefix-sum queries by probing $q$ cells if there is a map $Enc : \{0,1\}^n \to [n]^u$, $n$ sets $Q(1), \ldots, Q(n) \subseteq [u]$ of size $q$ each and $n$ decoding functions $d_1, \ldots, d_n$ mapping $[n]^q$ to $[n]$ such that for every $x \in \{0,1\}^n$ and every $i \in [n]$:*

$$Sum(i) := \sum_{k \leq i} x_k = d_i\left(Enc(x)|_{Q(i)}\right),$$

*where $Enc(x)|_{Q(i)}$ denotes the $q$ cells of size $n$ of $Enc(x) \in [n]^u$ indexed by $Q(i)$.*

**Theorem 1.1** (Lower bound for prefix sums). (Restated.) *To store $\{0,1\}^n$ in $[n]^u$ so that each $Sum(i) := \sum_{k \leq i} x_k$ query can be computed by non-adaptively probing $q$ cells of $\lg_2 n$ bits, one needs memory*

$$u \cdot \lg_2 n \geq n - 1 + n/\lg^{A \cdot q} n,$$

*where $A$ is an absolute constant.*

The proof relies on a few lemmas which we describe next.

The first is the separator lemma which shows that given any family of small subsets of some universe, we can remove a few $w/g$ elements from the universe to find many $w$ disjoint sets in our family. (The sets are disjoint if no element is contained in any two of them; the empty set is disjoint from anything else.)

**Lemma 2.3** (Separator). *For every $n$ sets $Q(1), Q(2), \ldots, Q(n)$ of size $q$ each and every desired "gap" $g$, there is $w \in [n/(g \cdot q)^q, n]$ and a set $B$ of size $|B| \leq w/g$ such that there are $\geq w$ disjoint sets among*

$$Q(1) \setminus B, Q(2) \setminus B, \ldots, Q(n) \setminus B.$$

The next lemma shows that given a set $V$ of $w$ indices in $[n]$ we can find a large subset of indices $V' \subseteq V$ that partition $[n]$ in intervals such that any interval starting at an even-indexed $v'$ is $\geq c$ times as large as the next one:

$$0 \underline{\qquad} v'_1 \underline{\,} v'_2 \underline{\qquad} v'_3 \underline{\,} v'_4 \underline{\quad} \cdots \underline{\quad} v'_{2k} \underline{\qquad} v'_{2k+1} \underline{\,} v'_{2k+2} \cdots$$

**Lemma 2.4** (Stretcher). *Let $1 \le v_1 < v_2 < \ldots < v_w \le n$ be $w$ indices in $[n]$. Let $c > 1$ and $n$ sufficiently large. Then there are $w' = 2\lfloor w/(c \cdot \lg n) \rfloor$ indices $V' := \{v'_1, v'_2, \ldots, v'_{w'}\} \subseteq \{v_1, v_2, \ldots, v_w\}$, ordered as $v'_1 < v'_2 < \ldots < v'_{w'}$, such that*

$$v'_{2k+1} - v'_{2k} \ge c(v'_{2k+2} - v'_{2k+1})$$

*for every $k = 0, 1, \ldots, w'/2 - 1$, where $v'_0 := 0$.*

For the next lemmas recall the concept of *entropy* $H$ of a random variable $X$, defined as $H(X) := \sum_x \Pr[X = x] \cdot \lg(1/\Pr[X = x])$ and conditional entropy $H(X|Y) := E_{y \in Y} H(X|Y = y)$ (cf. [CT06, Chapter 2]).

The following is the information-theoretic lemma showing that if one conditions uniformly distributed random variables $X_1, \ldots, X_n$ on an event that happens with noticeable probability, then even following the conditioning most groups of $q$ variables are close to being uniformly distributed. See [Vio09] and the references therein.

**Lemma 2.5** (Information-theoretic). *Let $X = (X_1, \ldots, X_n)$ be a collection of independent random variables where each one of them is uniformly distributed over a set $S$. Let $A \subseteq S^n$ be an event such that $\Pr[X \in A] \ge 2^{-a}$, and denote by $(X'_1, \ldots, X'_n)$ the random variables conditioned on the event $A$. Then for any $\eta > 0$ and integer $q$ there exists a set $G \subseteq [n]$ such that $|G| \ge n - 16 \cdot q \cdot a/\eta^2$ and for any $q$ indices $i_1 < i_2 < \ldots < i_q \in G$ we have the distribution $(X'_{i_1}, X'_{i_2}, \ldots, X'_{i_q})$ is $\eta$-close to uniform.*

We need a variant of the above lemma where the variables keep high entropy even when conditioning on the ones before.

**Lemma 2.6** (Information-theoretic II). *Let $Z$ be uniformly distributed in a set $X \subseteq \{0,1\}^n$ of size $|X| = 2^{n-a}$. Let $Z = (Z_1, \ldots, Z_k)$ where $Z_i \in \{0,1\}^{s_i}$ (so that $\sum_{i \le k} s_i = n$). There is a set $G \subseteq [k]$ of size $|G| \ge k - a/\epsilon$ such that for any $i \in G$ we have*

$$H(Z_i | Z_1 Z_2 \ldots Z_{i-1}) \ge s_i - \epsilon.$$

*In particular, $Z_i$ is $4\sqrt{\epsilon}$ close to uniform over $\{0,1\}^{s_i}$.*

Finally, the next lemma lets us turn high entropy of a block of variables conditioned on the previous ones into bounds on the probabilities (2) and (4) in the overview Section 1.1.

**Lemma 2.7** (Entropy-sum). *Let $X_1, X_2, \ldots, X_n$ be $0-1$ random variables, and $p < i < j$ three indices in $[n]$ such that for $\ell := (i-p)$ and $d := j-i$ we have $\ell \ge c \cdot d$ for a sufficiently large $c$. Suppose that*

$$H(X_{p+1}, X_{p+2}, \ldots, X_j | X_1, X_2, \ldots, X_p) \ge \ell + d - 1/c.$$

*Then there exists an integer t such that*

$$\Pr_X\left[\sum_{k\leq j} X_k \geq t + \ell/2 + d/2 + c^{1/3}\sqrt{d}\right] \geq 1/10, \text{ and}$$

$$\Pr_X\left[\sum_{k\leq i} X_k < t + \ell/2\right] \geq 1/10, \text{ but}$$

$$\Pr_X\left[\sum_{k\leq j} X_k \geq t + \ell/2 + d/2 + c^{1/3}\sqrt{d} \bigwedge \sum_{k\leq i} X_k < t + \ell/2\right] \leq 1/1000 (\ll 1/10 \cdot 1/10).$$

## 2.1 Proof of lower bound

Let $c$ be a fixed, sufficiently large constant to be determined later, and let $n$ go to infinity. We prove the theorem for $A := c + 1$: we assume there exists a representation with redundancy $n/\lg^{A\cdot q} n - 1$ and derive a contradiction. First, we assume $q \geq 1$ for else the theorem is trivially true. We further assume that $q \leq (\log n)/2(c+1)\lg \lg n$ for else the redundancy is $< 0$ and again the theorem trivially true.

*Separator:* We apply Lemma 2.3 to the sets $Q(1), \ldots, Q(n)$ with gap $g := \lg^c n$ to obtain $w \in [n/(q \cdot \lg^c n)^q, n]$ and a set $B \subseteq [u]$ of size $|B| \leq w/\lg^c n$ such that there are $\geq w$ disjoint sets among

$$Q(1) \setminus B, Q(2) \setminus B, \ldots, Q(n) \setminus B.$$

Let these sets be

$$Q(v_1) \setminus B, Q(v_2) \setminus B, \ldots, Q(v_w) \setminus B,$$

and let $V := \{v_1, v_2, \ldots, v_w\} \subseteq [n]$ be the corresponding set of indices. Observe that $w \geq n/(q \cdot \lg^c n)^q \geq n/(\lg^{c+1} n)^{(\lg n)/2(c+1)\lg \lg n} \geq \sqrt{n}$.

Over the choice of a uniform input $x \in \{0,1\}^n$, consider the most likely value $z$ for the $w/\lg^c n$ cells indexed by $B$. Let us fix this value for the cells. Since this is the most likely value, we are still decoding correctly a set $X$ of $2^n/n^{|B|}$ inputs. From now on we focus on this set of inputs. Since these values are fixed, we can modify our decoding as follows. For every $i$ define $Q'(i) := Q(i) \setminus B$ and also let $d'_i$ be $d_i$ where the values of the probes corresponding to cells in $B$ have been fixed to the corresponding value in $z$. By renaming variables, letting $u' := u - |B|$ and $Enc' : \{0,1\}^n \to [n]^{u'}$ be $Enc$ restricted to the cells in $[u] \setminus B$, we see that we are now encoding $X$ in $[n]^{u'}$ in the following sense: for every $x \in X$ and every $i \in [n]$:

$$\sum_{k\leq i} x_k = d'_i \left(Enc'(x)|_{Q'_i}\right), \tag{6}$$

where note for any $i, j \in V$ we have $Q'(i) \bigcap Q'(j) = \emptyset$.

*Uniform cells:* To the choice of a uniform $x \in X \subseteq \{0,1\}^n$ there corresponds a uniform encoding $y \in Y \subseteq [n]^{u'}$, where

$$|X| = |Y| \geq 2^n/n^{|B|} = 2^{n - w/\lg^{c-1} n}. \tag{7}$$

Let $y = (y_1, \ldots, y_{u'})$ be selected uniformly in $Y \subseteq [n]^{u'}$. By Lemma 2.5 there is a set $G \subseteq [u']$ of size

$$|G| \geq u' - 16 \cdot 2q \cdot \lg(n^{u'}/|Y|) \cdot c^2 \geq u' - 16 \cdot 2q \cdot \lg\left(\frac{n^u/n^{|B|}}{2^n/n^{|B|}}\right) \cdot c^2$$

$$= u' - 32q \cdot r \cdot c^2,$$

where $r := (u \lg n) - n = n/\lg^{A \cdot q} n - 1$ is the redundancy of the data structure, such that for any $2q$ indices $k_1, k_2, \ldots, k_{2q}$ the cells $y_{k_1}, \ldots, y_{k_{2q}}$ are jointly $(1/c)$-close to uniform. Since the sets $Q'(v_1), Q'(v_2), \ldots, Q'(v_w) \subseteq [u']$ are disjoint, there is a set $V_2 \subseteq V$ such that for any $i, j \in V_2$ and $y$ uniform in $Y$ the distribution

$$(y|_{Q'(i)}, y|_{Q'(j)})_{y \in Y}, \text{ is } 1/c \text{ close to uniform over } [n]^{|Q'(i)|+|Q'(j)|}, \tag{8}$$

and the size of $|V_2|$ is

$$|V_2| \geq w - 32q \cdot r \cdot c^2 \geq w - 32q \cdot \frac{n}{\lg^{A \cdot q} n} \cdot c^2 = w - 32q \cdot \frac{n}{\lg^{(c+1) \cdot q} n} \cdot c^2 \geq w/2, \tag{9}$$

where the last inequality (9) holds because $w \geq n/(q \cdot \lg^c n)^q$. Specifically, the inequality is implied by $((\lg n)/q)^q \leq 64q \cdot c^2$, which is true because $q \leq (\lg n)/(2(c+1) \lg \lg n)$.

*Stretcher:* Apply Lemma 2.4 to $V_2$ to obtain a subset $V_3 \subseteq V_2$ of even size

$$w' := |V_3| \geq 2\lfloor |V_2|/(c \cdot \lg n) \rfloor \geq 2\lfloor w/(2c \cdot \lg n) \rfloor \geq w/(2c \cdot \lg n), \tag{10}$$

such that if $v'_1 < v'_2 < \ldots < v'_{|V_3|}$ is an ordering of the elements of $V_3$ we have

$$v'_{2k+1} - v'_{2k} \geq c(v'_{2k+2} - v'_{2k+1}) \tag{11}$$

for every $k = 0, 1, \ldots, w'/2 - 1$.

*Entropy in input bits.* For a uniform $x \in X$ consider the $w'/2$ random variables $Z_k$ where for $0 \leq k < w'/2$ $Z_k$ stands for the $s_k$ bits of $x$ from position $v'_{2k} + 1$ to $v'_{2k+2}$,

$$Z_k := x_{v'_{2k}+1} x_{v'_{2k}+2} \ldots x_{v'_{2k+2}} \in \{0,1\}^{s_k},$$

and $Z_{w'/2-1}$ is padded to include the remaining bits as well, $Z_{w'/2-1} = x_{v'_{w'-2}+1} x_{v'_{w'-2}+2} \ldots x_{v'_{w'}} \ldots x_n$:

$$\underbrace{x_1 \text{------} x_{v'_1} \text{---} x_{v'_2}}_{Z_0} \underbrace{x_{v'_2+1} \text{------} x_{v'_3} \text{---} x_{v'_4}}_{Z_1} \underbrace{x_{v'_4+1} \text{------} x_{v'_5} \text{---} x_{v'_6}}_{Z_2} \ldots$$

Recalling the bound (7) on the size of $|X|$, apply Lemma 2.6 to conclude

$$H(Z_k | Z_1 Z_2 \ldots Z_{k-1}) \geq s_k - 1/c \tag{12}$$

for

$$w'/2 - c \cdot w/\lg^{c-1} n \geq w/(4c \cdot \lg n) - c \cdot w/\lg^{c-1} n \geq 1 \tag{13}$$

variables $Z_k$. Fix an index $k$ such that Equation (12) holds for $Z_k$, and let

$$p := v'_{2k}, \qquad i := v'_{2k+1}, \qquad j := v'_{2k+2}$$

be the corresponding indices, where $p$ is either 0 or in $V_3$ and $\{i, j\} \subseteq V_3$. We can rewrite Equation (12) as $H(x_{p+1}x_{p+2}\ldots x_j | x_1 x_2 \ldots x_p)$. Using in addition Equation (11) we are in the position to apply Lemma 2.7 ($\ell := (i - p) \geq c \cdot d := c \cdot (j - i)$). Let $t$ be the integer in the conclusion of Lemma 2.7, and let

$$s := t + (\ell + d)/2 + c^{1/3}\sqrt{d}, \qquad s' := t + \ell/2.$$

Let $U'$ denote the uniform distribution over the $u'$ cells, i.e., over $[n]^{u'}$. We have the following contradiction:

$$\Pr_{x \in X}\left[\sum_{k \leq j} x_k \geq s \wedge \sum_{k \leq i} x_k < s'\right]$$

$$= \Pr_{y \in Y}\left[d'_j(y|_{Q'_j}) \geq s \wedge d'_i(y|_{Q'_i}) < s'\right] \qquad \text{(By Equation (6))}$$

$$\geq \Pr_{U'}\left[d'_j(U'|_{Q'_j}) \geq s \wedge d'_i(U'|_{Q'_i}) < s'\right] - 1/c \qquad \text{(By (8))}$$

$$= \Pr_{U'}\left[d'_j(U'|_{Q'_j}) \geq s\right] \cdot \Pr_{U'}\left[d'_i(U'|_{Q'_i}) < s'\right] - 1/c \qquad \text{(Because } Q'(i) \cap Q'(j) = \emptyset\text{)}$$

$$\geq \left(\Pr_{y \in Y}\left[d'_j(y|_{Q'_j}) \geq s\right] - 1/c\right)\left(\Pr_{y \in Y}\left[d'_i(y|_{Q'_i}) < s'\right] - 1/c\right) - 1/c \qquad \text{(By (8) again)}$$

$$= \left(\Pr_{x \in X}\left[\sum_{k \leq j} x_k \geq s\right] - 1/c\right)\left(\Pr_{x \in X}\left[\sum_{k \leq i} x_k < s'\right] - 1/c\right) - 1/c \qquad \text{(By Equation (6) again)}$$

$$\geq (1/10 - 1/c)(1/10 - 1/c) - 1/c \qquad \text{(By Lemma 2.7)}$$

$$> 1/200 \qquad \text{(For large enough } c\text{)}$$

which contradicts Lemma 2.7.

## 3 Lemmas

In this section we restate and prove the lemmas needed for the proof of our main theorem.

**Lemma 2.3** (Separator). (Restated.) *For every $n$ sets $Q(1), Q(2), \ldots, Q(n)$ of size $q$ each and every desired "gap" $g$, there is $w \in [n/(g \cdot q)^q, n]$ and a set $B$ of size $|B| \leq w/g$ such that there are $\geq w$ disjoint sets among*

$$Q(1) \setminus B, Q(2) \setminus B, \ldots, Q(n) \setminus B.$$

*Proof.* Set $k_0 := n/(g \cdot q)^q$. Initialize $B := \emptyset$. Consider the following procedure with stages $i = 0, 1, \ldots, q$. We maintain the following invariants: (1) at the beginning of stage $i$ our family consists of sets of size $q - i$ and (2) at the beginning of any stage $i \geq 1$, $|B| \leq k_0 \cdot g^{i-1}q^i$.

*The i-th stage:* Consider the family $(Q(1) \setminus B, Q(2) \setminus B, \ldots, Q(n) \setminus B)$. If it contains $\geq k_0(g \cdot q)^i$ disjoint sets then we successfully terminate because by the invariant $|B| \leq k_0 \cdot g^{i-1} q^i$.

If not, there must exist a covering $C$ of size $k_0(g \cdot q)^i(q-i)$ of the family, i.e., a set that intersects every element in our family. To see this, greedily collect in a set $S$ as many disjoint sets from our family as possible. We know we will stop with $|S| < k_0(g \cdot q)^i$. This means that every set in our family intersects some of the sets in $S$. Since the sets in the family have size at most $(q-i)$, the set $C$ of elements contained in any of the sets in $S$ constitutes a covering and has size $|C| \leq k_0(g \cdot q)^i(q-i)$.

Let $B := B \bigcup C$. We now finish the stage. Note that we have reduced the size of our sets by 1, maintaining Invariant (1). To see that Invariant (2) is maintained, note that if $i = 0$ then $|B| = |C| \leq k_0 \cdot q$, as desired. Otherwise, for $i \geq 1$, note that by Invariant (2) and the bound on $|C|$ we have

$$|B| \leq |C| + k_0 \cdot g^{i-1} q^i \leq k_0(g \cdot q)^i(q-i) + k_0 \cdot g^{i-1} q^i \leq k_0 \cdot g^i \cdot q^{i+1},$$

and thus Invariant (2) is maintained.

To conclude, note that the procedure terminates at stage $q$ at most, for at stage $q$ our family consists of $n = k_0(g \cdot q)^q$ empty sets which are all disjoint. $\square$

**Lemma 2.4** (Stretcher). (Restated.) *Let $1 \leq v_1 < v_2 < \ldots < v_w \leq n$ be $w$ indices in $[n]$. Let $c > 1$ and $n$ sufficiently large. Then there are $w' = 2\lfloor w/(c \cdot \lg n) \rfloor$ indices $V' := \{v'_1, v'_2, \ldots, v'_{w'}\} \subseteq \{v_1, v_2, \ldots, v_w\}$, ordered as $v'_1 < v'_2 < \ldots < v'_{w'}$, such that*

$$v'_{2k+1} - v'_{2k} \geq c(v'_{2k+2} - v'_{2k+1})$$

*for every $k = 0, 1, \ldots, w'/2 - 1$, where $v'_0 := 0$.*

*Proof.* Set $s := 0, t := \lfloor c \cdot \lg n \rfloor$ and define $v_0 := 0$. While $s \leq w - t$, consider the first $i : 0 < i \leq t - 1$ for which

$$v_{s+i} - v_s \geq c(v_{s+i+1} - v_{s+i}). \tag{14}$$

Add $v_{s+i}, v_{s+i+1}$ to $V'$. Set $s := s + i + 1$ and repeat.

This gives $w' \geq 2\lfloor w/t \rfloor \geq 2\lfloor w/(c \cdot \lg n) \rfloor$ indices, as desired, assuming we can always find $i : 0 < i \leq t - 1$ for which (14) holds. Suppose not. We have the following contradiction:

$$v_{s+t} - v_s = v_{s+t} - v_{s+t-1} + v_{s+t-1} - v_s$$
$$> (1+1/c)(v_{s+t-1} - v_s) > (1+1/c)^2(v_{s+t-2} - v_s) > \ldots > (1+1/c)^{t-1}(v_{s+1} - v_s)$$
$$\geq (1+1/c)^{t-1} = (1+1/c)^{\lfloor c \cdot \lg n \rfloor - 1} \geq (1+1/c)^{c \cdot \lg n}/(1+1/c)^2 \geq (2.25)^{\lg n}/(1+1/c)^2 \gg n,$$

for $c > 1$ and sufficiently large $n$. $\square$

We list next a few standard properties of entropy that we will use in the proofs.

**Fact 1.** *Entropy satisfies the following.*

1. Chain rule: *For any random variables $X, Y$, and $Z$: $H(X, Y|Z) = H(X|Z) + H(Y|X, Z)$* [CT06, Equation 2.21].

2. Conditioning reduces entropy: *For any random variables $X, Y, Z$ we have $H(X|Y) \geq H(X|Y, Z)$* [CT06, Equations 2.60 and 2.92].

3. High entropy implies uniform: *Let $X$ be a random variable taking values in a set $S$ and suppose that $H(X) \geq \lg |S| - \alpha$; then $X$ is $4\sqrt{\alpha}$-close to uniform* [CK82, Chapter 3; Exercise 17].

**Lemma 2.6** (Information-theoretic II). (Restated.) *Let $Z$ be uniformly distributed in a set $X \subseteq \{0,1\}^n$ of size $|X| = 2^{n-a}$. Let $Z = (Z_1, \ldots, Z_k)$ where $Z_i \in \{0,1\}^{s_i}$ (so that $\sum_{i \leq k} s_i = n$). There is a set $G \subseteq [k]$ of size $|G| \geq k - a/\epsilon$ such that for any $i \in G$ we have*

$$H(Z_i | Z_1 Z_2 \ldots Z_{i-1}) \geq s_i - \epsilon.$$

*In particular, $Z_i$ is $4\sqrt{\epsilon}$ close to uniform over $\{0,1\}^{s_i}$.*

*Proof.* We have $H(Z) = \log |X| = n - a = \sum s_i - a$. By the chain rule for entropy,

$$\sum_{i \leq k} (s_i - H(Z_i | Z_1 Z_2 \ldots Z_{i-1})) = a.$$

Applying Markov inequality to the non-negative random variable $s_i - H(Z_i | Z_1 Z_2 \ldots Z_{i-1})$ (for random $i \in [k]$), we have

$$\Pr_{i \in [k]} [s_i - H(Z_i | Z_1 Z_2 \ldots Z_{i-1}) \geq \epsilon] \leq a/(k \cdot \epsilon),$$

yielding the desired $G$.

The "in particular" part is an application of Items (2) and (3) in Fact 1. □

**Lemma 2.7** (Entropy-sum). (Restated.) *Let $X_1, X_2, \ldots, X_n$ be $0-1$ random variables, and $p < i < j$ three indices in $[n]$ such that for $\ell := (i - p)$ and $d := j - i$ we have $\ell \geq c \cdot d$ for a sufficiently large $c$. Suppose that*

$$H(X_{p+1}, X_{p+2}, \ldots, X_j | X_1, X_2, \ldots, X_p) \geq \ell + d - 1/c.$$

*Then there exists an integer $t$ such that*

$$\Pr_X \left[ \sum_{k \leq j} X_k \geq t + \ell/2 + d/2 + c^{1/3}\sqrt{d} \right] \geq 1/10, \text{ and}$$

$$\Pr_X \left[ \sum_{k \leq i} X_k < t + \ell/2 \right] \geq 1/10, \text{ but}$$

$$\Pr_X \left[ \sum_{k \leq j} X_k \geq t + \ell/2 + d/2 + c^{1/3}\sqrt{d} \bigwedge \sum_{k \leq i} X_k < t + \ell/2 \right] \leq 1/1000 (\ll 1/10 \cdot 1/10).$$

*Proof.* Let us start with the last inequality, because we can prove it without getting our hands on $t$. First, by Item (3) in Fact 1, in particular the distribution of $X_{i+1}, X_{i+2}, \ldots, X_j$ is $4/\sqrt{c}$ close to the uniform $U_1 U_2 \ldots U_d$. We have

$$\Pr_X \left[ \sum_{k \leq j} X_k \geq t + \ell/2 + d/2 + c^{1/3}\sqrt{d} \bigwedge \sum_{k \leq i} X_k < t + \ell/2 \right]$$

$$\leq \Pr_X \left[ \sum_{k=i+1}^{j} X_k \geq d/2 + c^{1/3}\sqrt{d} \right] \leq \Pr_X \left[ \sum_{k=1}^{d} U_k \geq d/2 + c^{1/3}\sqrt{d} \right] + 4/\sqrt{c}$$

$$\leq 1/2000 + 4/\sqrt{c} \leq 1/1000,$$

where the second to last inequality follows from Chebyshev's inequality for sufficiently large $c$.

We now verify the first two inequalities in the conclusion of the lemma. Let $Y := X_1, X_2, \ldots, X_p$ stand for the prefix, and $Z := X_{p+1}, X_{p+2}, \ldots, X_j$ for the $\ell + d$ high-entropy variables. Let

$$A := \{y \in \{0,1\}^p : H(Z|Y=y) \geq \ell + d - 2/c\}$$

be the set of prefix values conditioned on which $Z$ has high entropy. We claim that $\Pr[Y \in A] \geq 1/2$. This is because, applying Markov Inequality to the non-negative random variable $\ell + d - H(Z|Y=y)$ (for $y$ chosen according to $Y$),

$$\Pr[Y \notin A] = \Pr_{y \in Y}[\ell + d - H(Z|Y=y) > 2/c] \leq$$

$$E_{y \in Y}[\ell + d - H(Z|Y=y)]/(2/c) = (\ell + d - H(Z|Y))/(2/c) \leq (1/c)/(2/c) = 1/2.$$

Note that for every $y \in A$ we have, by definition, that the $(\ell + d)$-bit random variable $(Z|Y=y)$ has entropy at least $\ell + d - 2/c$, and so by Item 3 in Fact 1 the random variable $(Z|Y=y)$ is ($\epsilon := 4\sqrt{2/c}$)-close to uniform over $\{0,1\}^{\ell+d}$. Therefore, for any subset $S \subseteq A$, the random variable

$$(Z|Y \in S) \text{ is } \epsilon\text{-close to uniform over } \{0,1\}^{\ell+d}. \tag{15}$$

Now define $t$ to be the largest integer such that

$$\Pr\left[Y \in A \wedge \sum_{k \leq p} Y_k \geq t\right] \geq 1/4. \tag{16}$$

Since by definition of $t$ we have $\Pr[Y \in A \wedge \sum_{k \leq p} Y_k \geq t+1] < 1/4$, we also have

$$\Pr\left[Y \in A \wedge \sum_{k \leq p} Y_k \leq t\right] \geq 1/2 - 1/4 = 1/4. \tag{17}$$

We obtain the desired conclusions as follows, denoting by $U_1, U_2, \ldots$, uniform and independent $0-1$ random variables. First,

$$\Pr\left[\sum_{k \leq j} X_k \geq t + (\ell + d)/2 + c^{1/3}\sqrt{d}\right]$$

$$\geq \Pr\left[\sum_{k \leq j} X_k \geq t + (\ell + d)/2 + \sqrt{\ell}/c^{1/6}\right] \quad \text{(Because } \ell \geq c \cdot d\text{)}$$

$$= \Pr\left[\sum_{k \leq p} Y_k + \sum_{k \leq \ell + d} Z_k \geq t + (\ell + d)/2 + \sqrt{\ell}/c^{1/6}\right]$$

$$\geq \Pr\left[\sum_{k \leq \ell + d} Z_k \geq (\ell + d)/2 + \sqrt{\ell}/c^{1/6} \bigg| Y \in A \wedge \sum_{k \leq p} Y_k \geq t\right] \cdot \Pr\left[Y \in A \wedge \sum_{k \leq p} Y_k \geq t\right]$$

$$\geq \left(\Pr\left[\sum_{k \leq \ell + d} U_k \geq (\ell + d)/2 + \sqrt{\ell}/c^{1/6}\right] - \epsilon\right)(1/4)$$

$$\geq \left(1/2 - \sqrt{\ell}/c^{1/6} \cdot \Theta(1/\sqrt{\ell}) - \epsilon\right)(1/4)$$

$$\geq \left(1/2 - \Theta(1/c^{1/6}) - \epsilon\right)(1/4) \geq 1/10,$$

where the third inequality uses (15) and (16), and and the fourth uses the standard estimate $\Pr\left[\sum_{k \leq \ell+d} U_k = (\ell+d)/2 + b\right] \leq \Pr\left[\sum_{k \leq \ell+d} U_k = \lfloor (\ell+d)/2 \rfloor\right] = \Theta(1/\sqrt{\ell+d}) \leq \Theta(1/\sqrt{\ell})$ (cf. [CT06, Lemma 17.5.1]).

Second,

$$\Pr\left[\sum_{k \leq i} X_k \leq t + \ell/2\right] = \Pr\left[\sum_{k \leq p} Y_k + \sum_{k \leq \ell} Z_k < t + \ell/2\right]$$

$$\geq \Pr\left[\sum_{k \leq \ell} Z_k < \ell/2 \bigg| Y \in A \wedge \sum_k Y_k \leq t\right] \cdot \Pr\left[Y \in A \wedge \sum_k Y_k \leq t\right]$$

$$\geq \left(\Pr\left[\sum_{k \leq \ell} U_k < \ell/2\right] - \epsilon\right) \cdot (1/4) \geq (1/2 - \epsilon) \cdot (1/4) \geq 1/10$$

for all sufficiently large $c$. Here the second inequality uses (15) and (17). □

## 4 Balanced brackets

In this section we prove our lower bound for balanced brackets. We start by formally defining the problem and then we restate our theorem.

**Definition 4.1** (Data structure for balanced brackets). *We say that we store* Bal := $\{x \in \{0,1\}^n : x$ *corresponds to a string of balanced brackets*$\}$ *in* $[n]^u$ *supporting match queries by probing $q$ cells if there is a map* Enc $: \{0,1\}^n \to [n]^u$, $n$ *sets* $Q(1), \ldots, Q(n) \subseteq [u]$ *of size $q$ each and $n$ decoding functions* $d_1, \ldots, d_n$ *mapping* $[n]^q$ *to* $[n]$ *such that for every* $x \in \{0,1\}^n$ *and every* $i \in [n]$:

$$\text{Match}(i) := \text{ index to bracket matching } i = d_i\left(\text{Enc}(x)|_{Q(i)}\right),$$

*where* $\text{Enc}(x)|_{Q(i)}$ *denotes the $q$ cells of size $n$ of* $\text{Enc}(x) \in [n]^u$ *indexed by* $Q(i)$.

**Theorem 1.2** (Lower bound for balanced brackets). (Restated.) *To store* Bal := $\{x \in \{0,1\}^n : x$ *corresponds to a string of balanced brackets*$\}$ *in* $[n]^u$, $n$ *even, so that each Match$(i)$ query can be computed by non-adaptively probing $q$ cells of* $\lg_2 n$ *bits, one needs memory*

$$u \cdot \lg_2 n \geq n - 1 + n/\lg^{A^q} n,$$

*where $A$ is an absolute constant.*

**Overview of the proof:** In the same spirit of the proof of Theorem 1.1, we show that a too efficient data structure allows us to break the dependencies between queries and gives the following contradiction, for some subset of inputs $X$ and indices $i < j$:

$$0 = \Pr\left[\text{Match}(i) > j \bigwedge \text{Match}(j) < i\right]$$
$$\approx \Pr[\text{Match}(i) > j] \cdot \Pr[\text{Match}(j) < i]$$
$$\geq \epsilon \cdot \epsilon.$$

Above, the first equality obviously holds because $i < j$. The next breaking of the dependencies is again obtained with the separator lemma plus the information-theoretic lemma. For the final inequalities we find two indices $i < j$ that are close to each other, and also make sure that the input bits between $i$ and $j$ have high entropy, and then we use standard estimates that bound these probabilities by $\epsilon := \Omega(1/\sqrt{j-i})$. Whereas the corresponding probabilities in the proof of Theorem 1.1 can be bounded from below by a constant, here the bound deteriorates with the distance of the indices. This forces us to use the separator lemma with different parameters and overall yields a weaker bound.

We need the following version of the separator lemma.

**Lemma 4.3** (Separator). *Let $c \geq 4$ be any fixed, given constant. For all sufficiently large $n$, for all $n$ sets $Q(1), Q(2), \ldots, Q(n)$ of size $q \leq (\lg \lg n)/c$ each, there are two integers $a, b \geq 1$ such that*

$$c \cdot a \leq b \leq c \cdot (2c)^q,$$

*and $n/\lg^b n \geq 1$, and there is a set $B$ of size $|B| \leq n/\lg^b n$ such that there are $\geq n/\lg^a n$ disjoint sets among*

$$Q(1) \setminus B, Q(2) \setminus B, \ldots, Q(n) \setminus B.$$

*Proof.* Let $d := 2c$, $L := \lg n$. Initialize $B := \emptyset$. Consider the following procedure with stages $i = 0, 1, \ldots, q$. We maintain two invariants: (1) at the beginning of stage $i$ the family $Q(1) \setminus B, Q(2) \setminus B, \ldots, Q(n) \setminus B$ consists of sets of size $q - i$ and (2) at the beginning of any stage $i \geq 1$,
$$|B| \leq n/L^{c \cdot d^{q-i}},$$
while at the beginning of stage $i = 0$ we have $|B| = 0$.

*The $i$-th stage:* Consider the family $(Q(1) \setminus B, Q(2) \setminus B, \ldots, Q(n) \setminus B)$. If it contains
$$n/L^{d^{q-i}} =: n/L^a$$
disjoint sets then we successfully terminate the procedure, because by the invariant $|B| \leq n/L^{c \cdot d^{q-i}} =: n/L^b$.

If not, there must exist a covering $C$ of size $(q - i) \cdot n/L^{d^{q-i}}$ of the family, i.e., a set that intersects every element in our family. To see this, greedily collect in a set $S$ as many disjoint sets from our family as possible. We know we will stop with $|S| \leq n/L^{d^{q-i}}$. This means that every set in our family intersects some of the sets in $S$. Since the sets in the family have size at most $(q - i)$, the set $C$ of elements contained in any of the sets in $S$ constitutes a covering and has size $|C| \leq (q - i) \cdot n/L^{d^{q-i}}$.

Let $B := B \bigcup C$. We now finish the stage. Note that we have reduced the size of our sets by 1, maintaining Invariant (1). To see that Invariant (2) is maintained, note that by Invariant (2) and the bound on $|C|$ we have
$$|B| + |C| \leq i \cdot n/L^{c \cdot d^{q-i}} + (q-i) \cdot n/L^{d^{q-i}} = q \cdot n/L^{d^{q-i}} \leq n/L^{c \cdot d^{q-i-1}},$$
where the last inequality holds because it is equivalent to
$$q \cdot L^{c \cdot d^{q-i-1}} \leq L^{d^{q-i}}$$
$$\Leftarrow \lg q + c \cdot d^{q-i-1} \lg \lg n \leq d^{q-i} \lg \lg n$$
$$\Leftarrow (\lg q)/\lg \lg n \leq d^{q-i} - c \cdot d^{q-i-1} = d^{q-i} - d^{q-i}/2 = d^{q-i}/2$$
and $d^{q-i}/2 \geq 1/2$ and since $q \leq (\lg \lg n)/c$ we have $(\lg q)/\lg \lg n \leq 1/2$ for large enough $n$.

Note that the procedure successfully terminates at some stage $i \leq q$ at most, for at stage $i = q$ our family consists of
$$n \geq n/L^{d^{q-q}} = n/L$$
empty sets which are all disjoint.

To conclude, it only remains to verify the bound $n/L^b \geq 1 \Leftrightarrow \lg n \geq b \lg \lg n$. Observe that $b = c \cdot d^{q-i}$ for some $i = 0, 1, \ldots, q$. Therefore
$$b \leq c \cdot d^q \leq c(2c)^{(\lg \lg n)/c} \leq c(\lg n)^{(\lg 2c)/c} \leq c(\lg n)^{3/4} \Rightarrow b \lg \lg n \leq c(\lg n)^{3/4} \lg \lg n \leq \lg n$$
where we use that $c \geq 4$ is fixed, and that $n$ is sufficiently large. $\square$

We then recall a few standard facts about balanced brackets.

**Lemma 4.4.** *The number of strings of length n that correspond to balanced brackets is $\binom{n}{n/2}/(n/2+1)$, if n is even.*

*Let $x = (x_1, x_2, \ldots, x_d)$ be uniformly distributed in $\{0,1\}^d$. Then*

$$\Pr[x_1 \text{ is an open bracket and is not matched by } x_2, x_3, \ldots, x_d]$$
$$= \Pr[x_d \text{ is a closed bracket and is not matched by } x_1, \ldots, x_{d-2}, x_{d-1}] \geq \alpha/\sqrt{d},$$

*for a universal constant $\alpha > 0$.*

*Remarks on the proof.* The first equality is the well-known expression for Catalan numbers. We now consider the second claim in the statement of the lemma. This probability is easily seen to be at least $1/2$ times the probability that a $+1, -1$ walk of length $d-1$ starting at 0 never falls below 0. Assuming without loss of generality that $d-1$ is even, the latter probability equals the probability that a $+1, -1$ walk of length $d-1$ starting at 0 ends at 0 (see, e.g., (2) in www.math.harvard.edu/~lauren/154/Outline14.pdf). Standard estimates (cf. [CT06, Lemma 17.5.1]) show this is $\geq \Theta(1/\sqrt{d})$. □

## 4.1 Proof of Theorem 1.2

Let $c$ be a fixed, sufficiently large constant to be determined later, and let $n$ go to infinity. We prove the theorem for $A := 2^c$. Specifically, we assume for the sake of contradiction that there exists a representation with redundancy $n/\lg^{A^q} n - 1$ and we derive a contradiction. We clearly must have $q \geq 1$. Also, note that we can assume that $q \leq (\lg \lg n)/c$, for else the redundancy is $-1 + n/\lg^{A^q} n \leq -1 + n/\lg^{\lg n} n < 0$, which is impossible. Then, since $q \leq (\lg \lg n)/c$, we can apply Lemma 4.3 to the sets $Q(1), \ldots, Q(n)$ to obtain integers $a, b \geq 1$ such that

$$c \cdot a \leq b \leq c \cdot (2c)^q, \tag{18}$$

$n/\lg^b n \geq 1$ and a set $B \subseteq [u]$ of size $|B| := n/\lg^b n$ such that there are at least $n/\lg^a n$ disjoint sets among

$$Q(1) \setminus B, Q(2) \setminus B, \ldots, Q(n) \setminus B.$$

Let these sets be

$$Q(v_1) \setminus B, Q(v_2) \setminus B, \ldots, Q(v_{n/\lg^a n}) \setminus B,$$

where we order $1 \leq v_1 \leq v_2 \leq \ldots \leq v_{n/\lg^a n} \leq n$, and let $V := \{v_1, v_2, \ldots, v_{n/\lg^a n}\}$ be the corresponding set of indices. Also define the parameter

$$d := 16 \lg^a n.$$

Over the choice of a uniform input $x \in \{0,1\}^n$, consider the most likely value $z$ for the $n/\lg^b n$ cells indexed by $B$. Let us fix this value for the cells. Since this is the most likely value, we are still decoding correctly a set $X$ of $|\text{Bal}|/n^{|B|}$ inputs. From now on we focus on this set of inputs. Since these values are fixed, we can modify our decoding as follows.

For every $i$ define $Q'(i) := Q(i) \setminus B$ and also let $d'_i$ be $d_i$ where the values of the probes corresponding to cells in $B$ have been fixed to the corresponding value in $z$. By renaming variables, letting $u' := u - |B|$ and $Enc' : \{0,1\}^n \to [n]^{u'}$ be $Enc$ restricted to the cells in $[u] \setminus B$, we see that we are now encoding $X$ in $[n]^{u'}$ in the following sense: for every $x \in X$ and every $i \in [n]$:

$$\text{Match}(i) = d'_i \left( Enc'(x)|_{Q'_i} \right), \tag{19}$$

where note for any $i, j \in V$ we have $Q'(i) \bigcap Q'(j) = \emptyset$.

*Uniform cells:* To the choice of a uniform $x \in X \subseteq \{0,1\}^n$ there corresponds a uniform encoding $y \in Y \subseteq [n]^{u'}$, where

$$|X| = |Y| \geq |\text{Bal}|/n^{|B|} = |\text{Bal}|/2^{n/\lg^{b-1} n}.$$

Let $y = (y_1, \ldots, y_{u'})$ be selected uniformly in $Y \subseteq [n]^{u'}$. By Lemma 2.5 with $\eta := 1/(c \cdot d) = 1/(c \cdot 16 \lg^a n)$ there is a set $G \subseteq [u']$ of size

$$|G| \geq u' - 16 \cdot 2q \cdot \lg(n^{u'}/|Y|) \cdot c^2 d^2 \geq u' - 16 \cdot 2q \cdot \lg\left(\frac{n^u/n^{|B|}}{|\text{Bal}|/n^{|B|}}\right) \cdot c^2 d^2$$

$$= u' - 32q \cdot r \cdot c^2 d^2,$$

where $r := (u \lg n) - \lg |\text{Bal}| \leq n/\lg^{A^q} n - 1$ is the redundancy of the data structure, such that for any $2q$ indices $k_1, k_2, \ldots, k_{2q}$ the cells $y_{k_1}, \ldots, y_{k_{2q}}$ are jointly $(1/(c \cdot d))$-close to uniform. Since the sets $Q'(v_1), Q'(v_2), \ldots, Q'(v_w) \subseteq [u']$ are disjoint, there is a set $V_2 \subseteq V$ such that for any $i, j \in V_2$ and $y$ uniform in $Y$ the distribution

$$(y|_{Q'(i)}, y|_{Q'(j)})_{y \in Y}, \text{ is } 1/(c \cdot d) = 1/(c \cdot 16 \lg^a n) \text{ close to uniform over } [n]^{|Q'(i)|+|Q'(j)|}, \tag{20}$$

and the size of $|V_2|$ is

$$|V_2| \geq |V| - 32q \cdot r \cdot c^2 \cdot d^2 \geq \frac{n}{\lg^a n} - 32q \cdot \frac{n}{\lg^{A^q} n} \cdot c^2 \cdot d^2 \geq \frac{n}{2 \lg^a n}, \tag{21}$$

where the last inequality (21) holds because, recalling $d = 16 \lg^a n$, it is implied by $\lg^{A^q - 3a} n \geq (16)^2 \cdot 64q \cdot c^2$ which is true because of the bounds on $q, c, n$, using that $A := 2^c, a \leq (2c)^q$.

*Close:* Order the indices in $V_2$ as $v'_1 < v'_2 < \ldots < v'_{n/(2 \lg^a n)}$. Consider the consecutive $\geq \lfloor |V_2|/2 \rfloor$ pairs $\{v'_1, v'_2\}, \{v'_3, v'_4\}, \ldots$. Throw away all those such that the distance of the corresponding indices is $\geq d$. Since $V_2 \subseteq [n]$, we throw away at most $n/d$ pairs. Put the indices of the remaining pairs in $V_3$. So $V_3$ contains at least

$$|V_3|/2 \geq \lfloor |V_2|/2 \rfloor - n/d \geq n/(8 \lg^a n) - n/(16 \lg^a n) \geq n/(16 \lg^a n) \tag{22}$$

of these pairs (and twice as many indices).

*Entropy in input bits:* For a uniform $x \in X$, let

$$x = Z_1 Z_2 \ldots Z_{|V_3|/2}$$

where the variables $Z_k$ are a partition of $x$ in consecutive bits such that each $Z_k$ contains exactly one pair from $V_3$. We now apply the information-theoretic Lemma 2.6 with $4\sqrt{\epsilon} := 1/(c\sqrt{d})$ (i.e., $\epsilon = 1/(16c^2 \cdot d)$) and using the bound on $|X|$

$$|X| \geq |\text{Bal}|/n^{|B|} \geq \frac{\binom{n}{n/2}}{(n/2+1)2^{n/\lg^{b-1} n}} \geq \Omega\left(\frac{2^n}{\sqrt{n}(n/2+1)2^{n/\lg^{b-1} n}}\right) \geq \frac{2^n}{n^2 \cdot 2^{n/\lg^{b-1} n}},$$

where we use Lemma 4.4 and that $n$ is sufficiently large, which implies

$$\lg(2^n/|X|) \leq 2\lg n + n/\lg^{b-1} n \leq n/\lg^{b-2} n$$

since $n/\lg^b n \geq 1$. The lemma guarantees that all but

$$16\frac{n}{\lg^{b-2} n} \cdot c^2 \cdot d = (16)^2 \frac{n}{\lg^{b-2-a} n} \cdot c^2 \tag{23}$$

variables $Z_k$ will be $1/(c\sqrt{d})$ close to uniform. Since by Equation (22) we have at least $|V_3|/2 \geq n/(16\lg^a n)$ variables $Z_k$, and $b \geq c \cdot a$, there exists a variable $Z_k$ that is $1/(c\sqrt{d})$ close to uniform. This variable contains one pair from $V_3$. Let $i < j \in [n]$ be the corresponding indexes, which recall satisfy $j - i \leq d$. Let $U'$ denote the uniform distribution on the $u'$ cells. We have the following contradiction:

$$0 = \Pr_{x \in X}\left[\text{Match}(i) > j \bigwedge \text{Match}(j) < i\right] \quad \text{(Because } i < j\text{)}$$

$$= \Pr_{y \in Y}\left[d_i'(y|_{Q_i'}) > j \bigwedge d_j'(y|_{Q_j'}) < i\right] \quad \text{(By Equation (19))}$$

$$\geq \Pr_{U'}\left[d_i'(U'|_{Q_i'}) > j \bigwedge d_j'(U'|_{Q_j'}) < i\right] - 1/(c \cdot d) \quad \text{(By (20))}$$

$$= \Pr_{U'}\left[d_i'(U'|_{Q_i'}) > j\right] \cdot \Pr_{U'}\left[d_j'(U'|_{Q_j'}) < i\right] - 1/(c \cdot d) \quad \text{(Because } Q'(i) \bigcap Q'(j) = \emptyset\text{)}$$

$$\geq \left(\Pr_{y \in Y}\left[d_i'(y|_{Q_i'}) > j\right] - 1/(c \cdot d)\right)\left(\Pr_{y \in Y}\left[d_j'(y|_{Q_j'}) < i\right] - 1/(c \cdot d)\right) - 1/(c \cdot d)$$

(By (20) again)

$$= \left(\Pr_{x \in X}[\text{Match}(i) > j] - 1/(c \cdot d)\right)\left(\Pr_{x \in X}[\text{Match}(j) < i] - 1/(c \cdot d)\right) - 1/(c \cdot d)$$

(By Equation (19) again)

$$\geq \left(\Pr_{x \in \{0,1\}^n}[\text{Match}(i) > j] - 2/(c \cdot \sqrt{d})\right)\left(\Pr_{x \in \{0,1\}^n}[\text{Match}(j) < i] - 2/(c \cdot \sqrt{d})\right) - 1/(c \cdot d)$$

(because $Z_k$ is $1/(c\sqrt{d})$ close to uniform)

$$\geq \left(\Omega(1/\sqrt{d}) - 2/(c \cdot \sqrt{d})\right)\left(\Omega(1/\sqrt{d}) - 2/(c \cdot \sqrt{d})\right) - 1/(c \cdot d) \quad \text{(By Lemma 4.4)}$$

$$> 0. \quad \text{(For large enough } c\text{)}$$

# 5   Open problems

One open problem is to handle adaptive probes. Another is to prove lower bounds for the membership problem: to our knowledge nothing is known even for two non-adaptive cell probes when the set size is a constant fraction of the universe. The difficulty in extending the results in this paper to the membership problem is that the correlations between query answers are less noticeable.

**Acknowledgments.**   We thank Mihai Pătraşcu for a discussion on the status of data structures for prefix sums.